\documentclass[a4paper,fleqn,usenatbib,hyperref]{mnras}
\usepackage{mathptmx}
\usepackage{threeparttable}
\usepackage[T1]{fontenc}
\usepackage{ae,aecompl}

\usepackage{graphicx}	
\usepackage{amsmath}	
\usepackage{amssymb}	
\usepackage{pdflscape}

\title[Dust in faint active asteroids]{Dust environment of
  active asteroids P/2019 A4
  (PANSTARRS) and P/2021 A5 (PANSTARRS)}  

\author[F. Moreno et al.]{
Fernando Moreno,$^{1}$\thanks{E-mail: fernando@iaa.es}
Javier Licandro,$^{2,3}$
Antonio Cabrera-Lavers,$^{2,4}$
\newauthor
David Morate$^{2,3}$
and Daniel Guirado$^{1}$\\ 
$^{1}$Instituto de Astrof\'\i sica de Andaluc\'\i a, CSIC, Glorieta de
la Astronom\'\i a s/n, 18008 Granada, Spain \\
$^{2}$Instituto de Astrof\'\i sica de Canarias, Spain \\
$^{3}$Departamento de Astrof\'{\i}sica, Universidad de La Laguna,
E-38206 La Laguna, Tenerife, Spain \\
$^{4}$GRANTECAN, Cuesta de San Jos\'e s/n, E-38712 Bre\~na Baja, La Palma, Spain
}

\date{Accepted XXX. Received YYY; in original form ZZZ}

\pubyear{2021}

\begin{document}
\label{firstpage}
\pagerange{\pageref{firstpage}--\pageref{lastpage}}
\maketitle

\begin{abstract}
We report on the characterisation of the dust activity and dynamical evolution of two
faint active asteroids, P/2019 A4, and P/2021 A5, observed with the
10.4m GTC using both imaging and spectroscopy.  
Asteroid P/2019 A4 activity is found to be
linked to an impulsive event occurring some $\pm$10 days
around perihelion, probably due to a 
collision or a rotational disruption. Its orbit is stable over 100 Myr timescales.
Dust tail models reveal a short-term 
burst producing (2.0$\pm$0.7)$\times$10$^6$
kg of dust for maximum particle radius $r_{max}$=1 cm. The spectrum of P/2019 A4 is featureless, and slightly redder than the Sun.  
P/2021 A5 was active 
 $\sim$50 days after perihelion, lasting $\sim$5 to $\sim$60
days, and ejecting (8$\pm$2)$\times$10$^6$ kg of dust 
for $r_{max}$=1 cm. The
orbital simulations show  
that a few percent of dynamical clones of P/2021 A5 are
unstable on 20-50 Myr timescales. Thus, P/2021 A5 might be an implanted
object from the JFC region or
beyond. These facts point to water ice sublimation as the activation
mechanism. This object also displays a featureless spectrum, but  
slightly bluer than the Sun. Nuclei sizes are estimated in 
the few hundred meters range for both asteroids. Particle ejection
speeds ($\approx$0.2 m s$^{-1}$) are consistent with escape
speeds from those small-sized objects.  

\end{abstract}




\section{Introduction}

Active asteroids constitute a new class of objects in the solar
system. They are characterised by being located in the main asteroid
belt, but, contrary to most objects in the belt, display cometary
appearance, i.e., dust comae and tails. Some 40 objects of this kind
have been discovered so far. The reasons of their activity are rather
diverse, including impact-induced, rotational break-up, thermal
fracture, or ice sublimation \citep{2015aste.book..221J}. While
    most of those objects seem native to the main belt, some of them have
    been shown to become unstable on timescales of a few tens of Myr
    \citep[e.g.][]{2009M&PS...44.1863H,2016Icar..277...19H}.  Given the
variety of phenomena that might lead to dust ejection, it is very 
convenient to increase the sample statistics to provide a better
knowledge of the physics involved. To that end, current sky surveys
such as the Panoramic Survey Telescope and  Rapid  Response  System
(PANSTARRS) and facilities to come such as the Vera Rubin
    Observatory are of the 
utmost importance in detecting these typically faint objects. 

Asteroid P/2019 A4 was discovered on UT 2019 Jan. 10.4 by PANSTARRS
\citep{2019CBET.4600....1W}. Its orbital elements $a$=2.614 au,
$e$=0.0896, and $i$=13.32$^\circ$ yield a Tisserand parameter 
respect to Jupiter of $T_J$=3.36, as most asteroids, yet displaying a
sizeable coma and tail. It is located in the middle portion of the belt,
as P/2010 A2 \citep[e.g.,][]{2010Natur.467..817J},
P/2016 G1 \citep[e.g.,][]{2016ApJ...826L..22M,2019A&A...628A..48H}, or
(6478) Gault \citep[e.g.,][]{2019A&A...624L..14M,2019ApJ...876L..19J},
but with a remarkably small eccentricity. In these three cases, the
activity has been found to be linked to short-term events, either a 
rotational mass loss or an impact as the possible causes of the
events. P/2021 A5 was also discovered by PANSTARRS \citep{MPEC2021A209}, on UT 2021 Jan. 06, when displaying a condensed coma and a
4$\arcsec$ tail. Its orbital elements
($a$=3.047 au, $e$=0.14, $i$=18.19$^\circ$) give $T_J$=3.147, and place
the object close to the outer main belt where ice sublimation
presumably starts to
become dominant over other mechanisms. Clear examples of this
mechanism at play are
324P/La Sagra \citep[e.g.,][]{2011ApJ...738L..16M,2015MNRAS.454L..81H,
2016AJ....152...77J} and 133P/Elst-Pizarro
\citep{2004AJ....127.2997H,2010MNRAS.403..363H}, that have shown
recurrent activity.

In this work, we present images and spectra of P/2019 A4 and P/2021 A5 obtained
with the Gran Telescopio Canarias (10.4m aperture) on the island of La
Palma. We study the dynamical evolution of the objects to shed
    some light on their origin and on the causes of their activity,
    and apply Monte Carlo dust tail models to the images obtained to
    determine the dust  
properties, and to place further constraints on the activation mechanisms.

\section{Orbital dynamics simulations}

To asses whether the two objects under study are native to the main
belt or interlopers coming from elsewhere, we propagated their orbits
backward in time up to 100 Myr. We remark that the short arc of
observations of these objects (in particular, P/2021 A5) results in
relatively large uncertainties in their best-fitted orbital parameters. In
any case, for each object, we integrated the
orbits of 200 dynamical clones according to the statistical
uncertainty of the 
current orbital elements. The orbital elements of those clones were 
generated using the covariance matrix 
\citep[e.g.][]{2010tod..book.....M}. The six-component vector of
orbital elements $\mathbf{x^\prime}$ 
of the dynamical clones are calculated according to the expression

\begin{equation}
  \mathbf{x^\prime} = \mathbf{A} \Psi + \mathbf{x}
\end{equation}

where $\mathbf{x}$ is the six-component vector of the nominal orbital
elements (best-fit solution), and $\mathbf{A}$ is a matrix verifying
$\mathbf{A} {\mathbf{A}}^\intercal=\mathbf{C} $, where $\mathbf{C}$ is
the covariance 
matrix. This matrix is obtained from the JPL Small-Body
Database for each asteroid, and matrix $\mathbf{A}$ is obtained from
$\mathbf{C}$ by a Cholesky decomposition using the
\texttt{FORTRAN} implementation as described in
\cite{1992nrfa.book.....P}. $\Psi$ is a six-dimensional vector whose 
components are normally distributed (Gaussian) deviates with zero mean and unit
variance, which are also obtained from the corresponding routine
described in \cite{1992nrfa.book.....P}.

\begin{figure}
  \begin{tabular}{c}
    \includegraphics[angle=-90,width=0.5\textwidth] {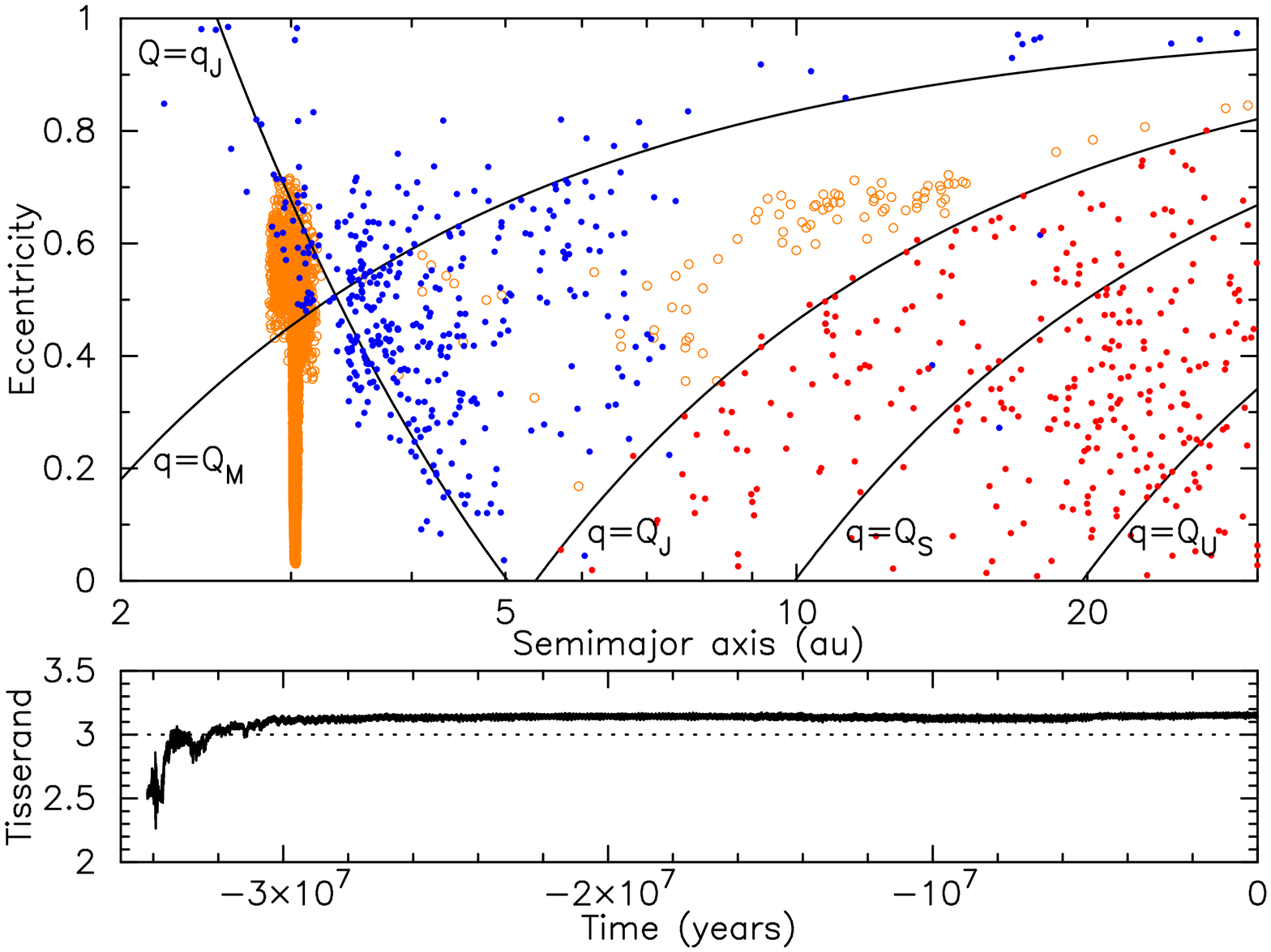} \\
    \includegraphics[angle=-90,width=0.5\textwidth] {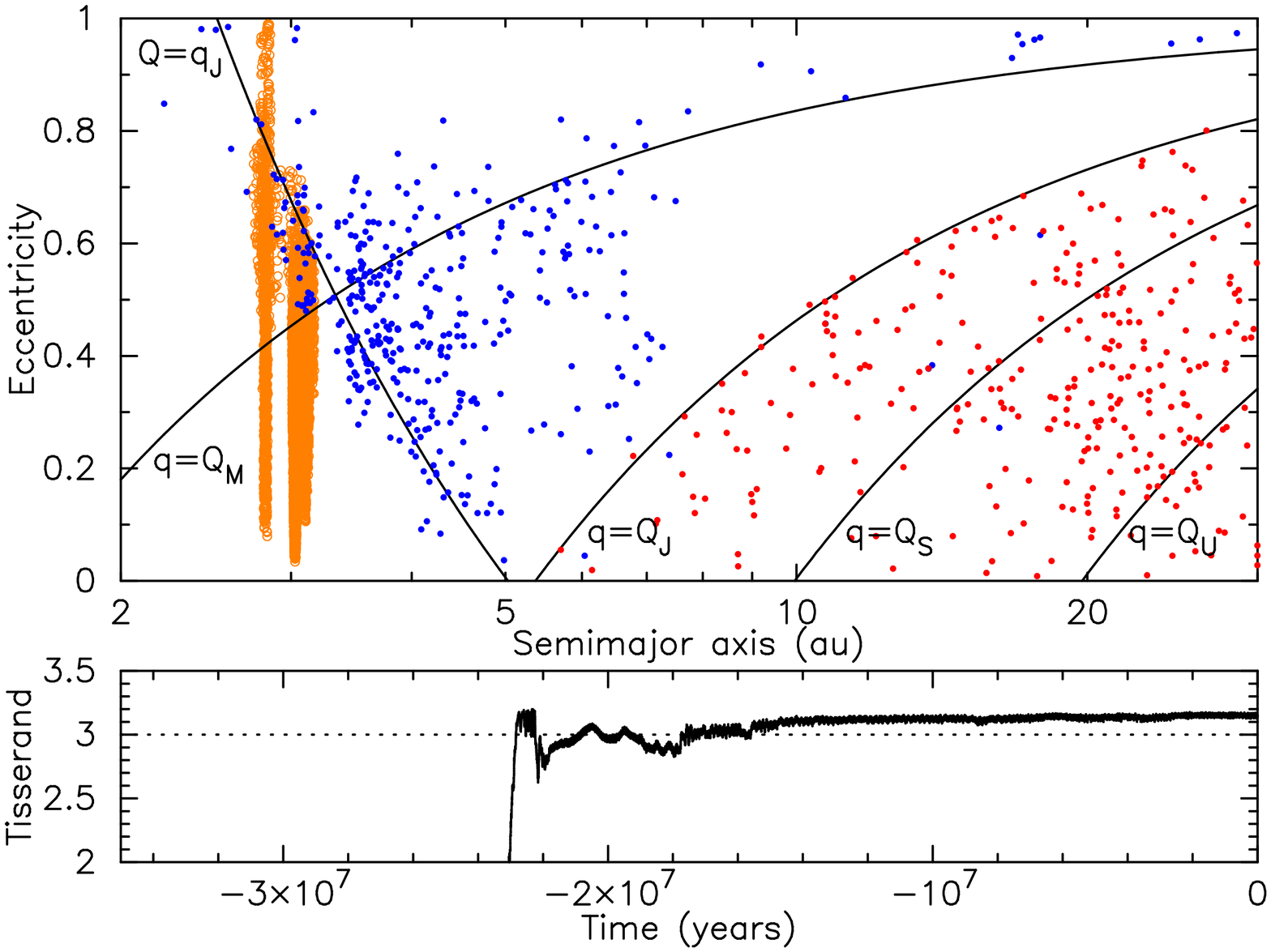} \\
    \includegraphics[angle=-90,width=0.5\textwidth] {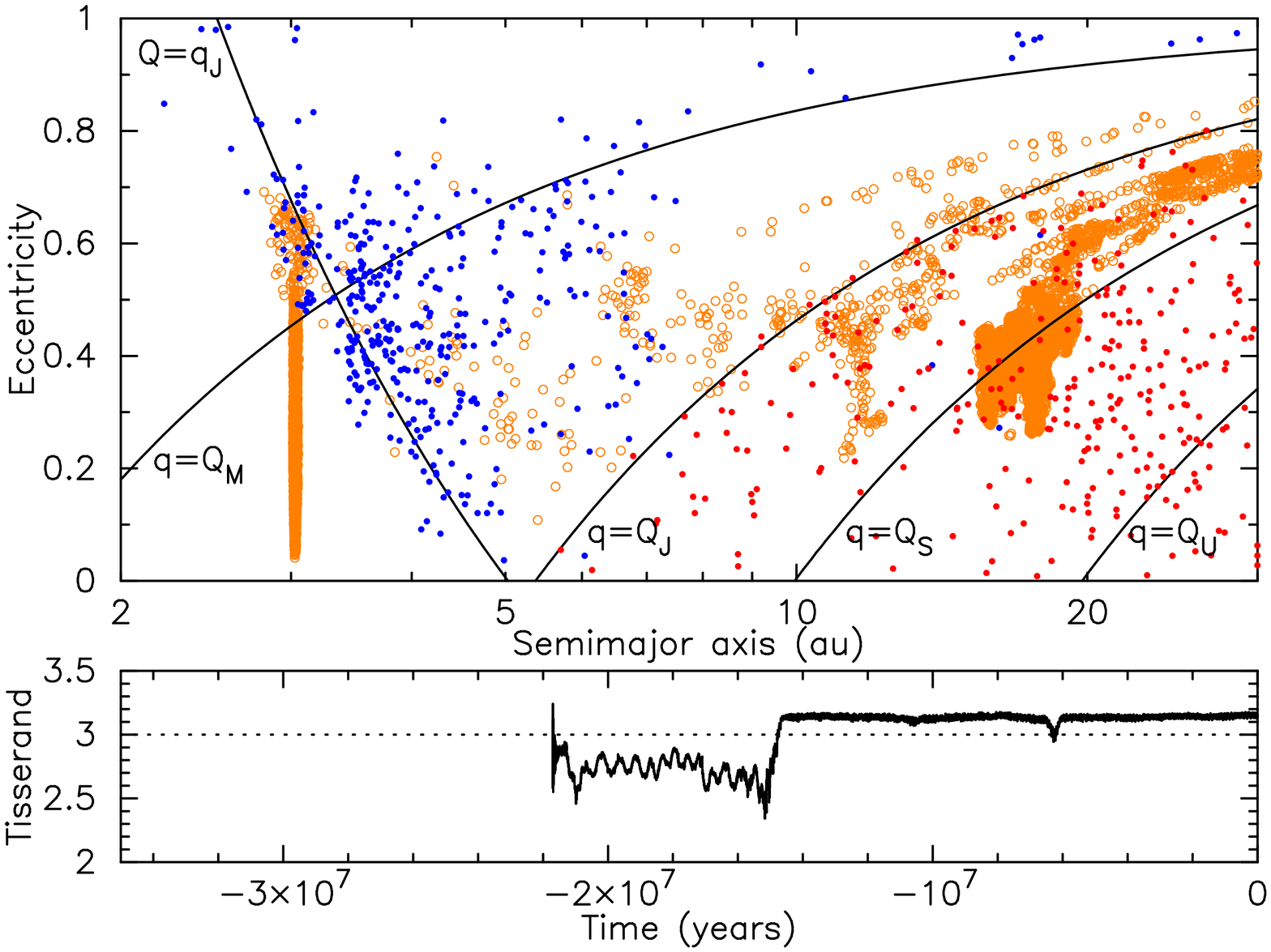} \\
    \end{tabular}
\caption{Semimajor axis versus eccentricity, and time evolution of the
  Tisserand parameter respect to Jupiter for some of the P/2021 A5
  unstable dynamical 
  clones. The clones are represented as orange open circles, the
  short-period comets as blue dots, and the subgroup of Centaurs
  having perihelia 
  5.2$<q<$30 au and semimajor axis $a<$30 as red dots. The solid
  curves show the loci of orbits having perihelia distances equal to 
  the aphelion distance of Mars ($Q_ M$), Jupiter ($Q_ J$), Saturn ($Q_S$), and 
  Uranus ($Q_ U$), and aphelion distances at Jupiter's
  perihelion ($q_J$). The dotted lines in the panels displaying the Tisserand
  parameters mark the classical distinction between asteroids $T>$3
  and comets $T<$3.}
   \label{dynamics}
\end{figure}

The time integrations were performed using the Bulirsch-St\"oer integrator
of \texttt{MERCURY} package \citep{1999MNRAS.304..793C}. All the eight
planets were included as major bodies, while all the dynamical clones
were considered as massless particles. Non-gravitational forces were
not included. We used an integration time step of 10 days. The orbits of 
all 200 dynamical clones generated for P/2019 A4 were stable over the 100 Myr
integration time, meaning that this object is very likely native from the
main belt. The case of P/2021 A5 is very different, however, because
of its proximity to the 9:4 Jupiter resonance (a=3.031 au). We have
noticed that 8 out of the 200 P/2021 A5 clones become eventually
unstable, being either ejected from the solar system or experience a
collision with some of the inner planets or the Sun on timescales of
order 20-50 Myr. Some of those ejected clones have intermediate
JFC-like 
and even Centaur-like orbits before ejection. Figure~\ref{dynamics}
displays intermediate orbital elements and the evolution of the
Tisserand parameter of some of those unstable
clones at 1000 year-intervals. The 
uppermost panel shows the typical dynamical evolution of the
unstable clones, where the 9:4 resonance induces excitation in eccentricity and
inclination, driving the particle to Mars and Jupiter crossing orbits,
and spending some time in JFC-like orbits before being ejected from the solar
system. Among those unstable clones, we have found two atypical cases in
the dynamical evolution. In one of those (central panels in
Figure~\ref{dynamics}), the particle migrates inwards,
experiencing several episodes of resonant oscillations, and being 
finally trapped in the strong 5:2 Jupiter resonance region at 2.825
au, from where 
its eccentricity is increased up to almost unity until a collision in
the inner solar system occurs. The 
other peculiar case (lowermost panels in Figure~\ref{dynamics})
correspond to a particle which, after being excited by the 9:4
resonance, spends some time in the JCF region, and then experiences several
resonant episodes in the Centaur region beyond Jupiter's orbit before
being finally ejected from the solar system.

While these calculations are not intended to provide robust
statistical results, as the number of test particles is not
sufficient, and the nominal orbital parameters have moderately large
uncertainties, they do indicate possible different origins of
these two active asteroids under analysis. As shown by the dynamical
experiments, P/2019 A4 is very likely a native member of the main belt,
while for P/2021 A5 there is some probability that it comes from
elsewhere, possibly from the JFC or Centaur regions, as a 
result from the dynamical evolution of the clone orbits. This 
suggests that P/2021 A5 might be an ice-bearing object. In this regard, we
 note that possible dynamical pathways from the JFC
region to the main belt have 
been previously explored by \cite{2002Icar..159..358F}, who did not
find any, and more recently by Hsieh and Haghighipour
\citep[see][and references
  therein]{2016Icar..277...19H}, who concluded that the
number of JFC-like interlopers in the main belt, albeit likely small,
might be non-zero.
    
\section{Observations}

\begin{figure*}
\centerline{\includegraphics[angle=-90,width=\textwidth] {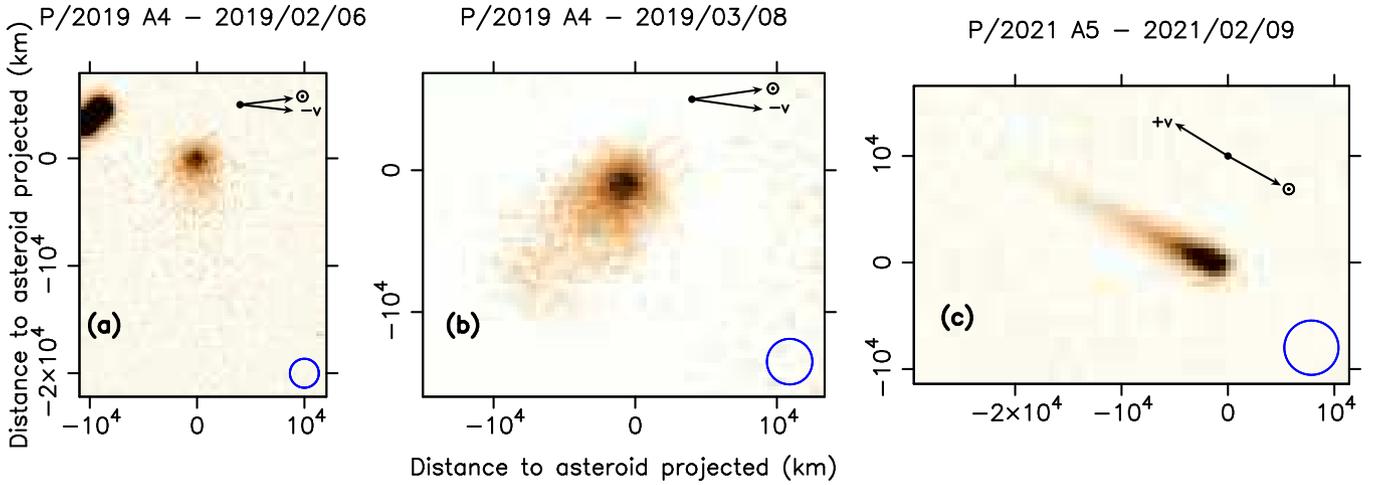}}
\caption{Median stack images of P/2019 A4 (panels (a) and (b))
  and P/2021 A5 (panel (c)) obtained with
  a $r'$ filter on the OSIRIS camera at the Gran Telescopio Canarias. 
 North is up, and East is to the left in all panels. The direction of 
 the Sun and the asteroid heliocentric velocity vector are
 indicated.  The blue circles near the bottom right corner in
     each image correspond
 to the size of the seeing \texttt{FWHM} for each image.}
   \label{images}
\end{figure*}

Images and spectra of asteroid P/2019 A4 have been obtained under
photometric conditions on 2019 February and 2019 March, and of P/2021
A5 on 2021 February. The log of observations is
shown in Table~\ref{logobs}. Images of P/2019
A4 were obtained 
on a CCD camera using a Sloan $r^\prime$ filter with the 
Optical System for Image and Low Resolution Integrated Spectroscopy
(OSIRIS) camera-spectrograph \citep{2010ASSP...14...15C} attached to
the 10.4m Gran Telescopio Canarias (GTC) at the Roque de los Muchachos
Observatory on the island of La Palma (Spain). For P/2021 A5,
additional images using the same setup were also obtained through $g^\prime$ and $i^\prime$
filters. The image scale was 
0.254$\arcsec$ pixel$^{-1}$. The images were bias-subtracted,
flat-fielded, and flux calibrated using standard procedures. Median stack
images were computed from the available frames on each night. The set
of images of these faint asteroids at the $r^\prime$ filter are displayed in Figure~\ref{images}. The
seeing was near 1.25$\arcsec$ \texttt{FWHM} during those nights, i.e., about 5
pixels \texttt{FWHM}.
\begin{table*}
	\centering
	\caption{Log of the observations. $R$ and $\Delta$ denote the asteroid
heliocentric and geocentric distances, respectively. $\chi$ is the
plane angle, i.e., the angle between the
observer and the asteroid orbital plane, and $\alpha$ is the phase angle.}
	\label{logobs}
	\begin{tabular}{lcccccccccc} 
		\hline
		Object & UT date & Days since     & $R$      & $\Delta$ &
                $\chi$  & $\alpha$ & Filter &  Grism &  Slit      &
                Exp. time \\
		           &               & Perihelion & (au) &(au)
                &(deg)    &(deg)        &          &
                &(arcsec) & (seconds)\\
		\hline
P/2019 A4 & 2019/02/06.85 & +63.4 & 2.389 & 1.487 & --2.72 & 12.87 & $r^\prime$ & --  & -- & 5x30\\
P/2019 A4 & 2019/02/06.87 & +63.4 & 2.389 & 1.487 & --2.72 & 12.87 & -- & R300R & 2.52 & 3x600\\
P/2019 A4 & 2019/03/08.86 & +93.4 & 2.399 & 1.770 & --5.53 & 21.46 & $r^\prime$ & -- &-- & 5x180\\ \hline
P/2021 A5 & 2021/02/09.84 & +90.7 & 2.645 & 2.812 & --0.86 & 20.54 & $g^\prime, r^\prime, i^\prime$ & -- &-- & 3x180 -- 9x180 -- 3x180\\           
P/2021 A5 & 2021/02/12.85 & +93.7 & 2.647 & 2.848 & --0.52 & 20.27 & -- & R300R & 1.2 & 3x600\\           
		\hline
	\end{tabular}
\end{table*}

\begin{table*}
	\centering
	\caption{Photometric results. Sloan magnitudes obtained using a 10 pixels diameter apertures are given
together with the lower limit of the absolute magnitude and the derived colours.}
	\label{magobs}
	\begin{tabular}{lccccccccc} 
		\hline
		Object & UT date & $g^\prime$ &  $r^\prime$ &  $i^\prime$ & $H_r$ & $(g^\prime- r^\prime)$ & $(r^\prime- i^\prime)$\\
		\hline
P/2019 A4 & 2019/02/06.85 & -- & $21.10 \pm 0.03$ & -- & $17.88 \pm 0.03$  &-- &--\\
P/2019 A4 & 2019/03/08.86 &-- & $22.42 \pm 0.05$  & -- & $18.51 \pm 0.05$  &-- &--\\ \hline
P/2021 A5 & 2021/02/09.84 & $22.17 \pm 0.05$  & $21.54 \pm 0.05$  & $21.48 \pm 0.05$  &$16.35 \pm 0.05$ &$0.63 \pm 0.05$ &$0.06 \pm 0.05$\\           
		\hline
	\end{tabular}
\end{table*}

\begin{figure}
  \centerline{\includegraphics[angle=-90,width=0.45\textwidth] {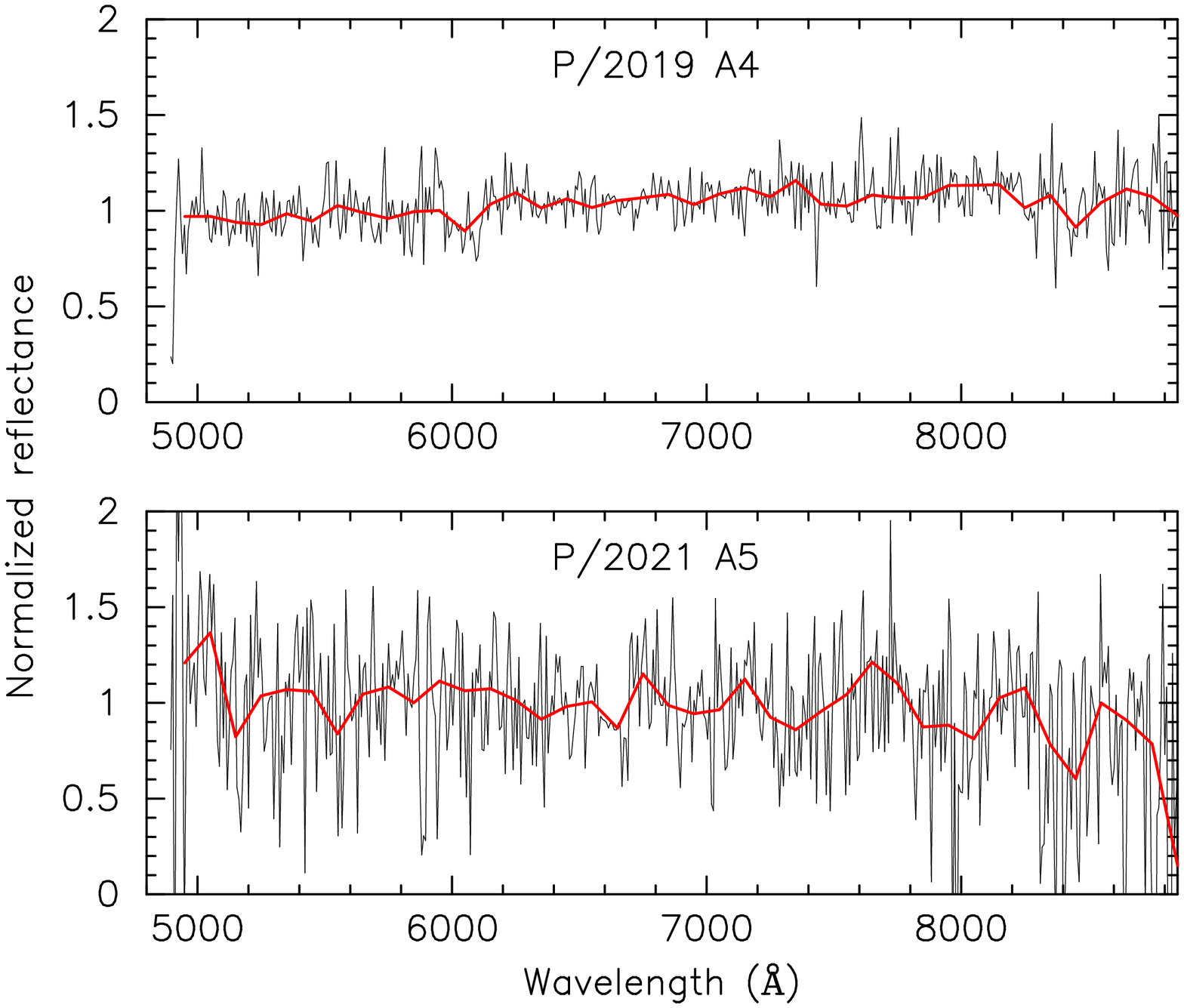}}
\caption{Upper panel: Normalized reflectance spectrum of P/2019 A4
  obtained with the 10.4m Gran Telescopio Canarias (black line). The
  thick red line is the median binned spectrum in 100 \AA ~resolution
  bins. Lower panel: same as the
  upper panel, but for asteroid P/2021 A5.}
   \label{spectrum}
\end{figure}

The spectral images were bias and flat-field corrected, using lamp
flats. Spectra were extracted and collapsed to one dimension, using
an aperture of $\pm$5 pixels centred at the maximum of the intensity
profile of the object. The wavelength calibration was performed using
Xe+Ne+HgAr lamps. Finally, the 3 spectra obtained of each comet were
averaged to obtain the final spectrum. To obtain the normalized
reflectance spectra of P/2019 A4 and P/2021 A5 shown in Figure~\ref{spectrum}
we observed two G2V stars that were used as solar
analogues. For P/2019 A4, we used the standard SA 98 978
    \citep{2009AJ....137.4186L} , while for
    P/2021 A5, we used SA 93 101 \citep{2000A&AS..146..169G}. The
    colours of those stars along 
    with the solar colours from \cite{2012ApJ...752....5R} are given in Table~\ref{stars}.

\begin{table}
      \centering
	\caption{Colour indices of the solar-type standard stars observed and
        the Sun.}
	\label{stars}
\begin{tabular}{lcccc} 
	  \hline
     Object & B-V & U-B & V-R & V-I  \\
          \hline
SA 93 101 & 0.647 & 0.154 & 0.352 & 0.691\\
SA 98 978 & 0.609 & 0.094 & 0.348 & 0.669 \\
      Sun & 0.653 & 0.166 & 0.352 & 0.702\\
	  \hline
\end{tabular}
\end{table}

    The reflectance spectra are featureless, with no  
evidence of CN, C$_2$ or C$_3$ emission bands typical of coma gas
emission nor the wide absorption band typical of the stony class
asteroids. Unfortunately, the SNR of both spectra is too low to attempt to
derive reliable gas production rates.  
The P/2019 A4 spectrum is  
slightly red with slope $S'=4 \pm 1 \%/100$ nm typical of a primitive
$X$-type asteroid. The P/2021 A5 spectrum is very noisy, and slightly
blue with a slope $S'= -3 \pm 1 \%/100$ nm as typical of a primitive $B$-type asteroid. From the computed $r^\prime$, $g^\prime$, and $i^\prime$
magnitudes (see Table \ref{magobs}), we get
$(g^\prime-r^\prime)$=0.63$\pm$0.05 and 
$(r^\prime-i^\prime)$=0.06$\pm$0.05.  The colours of the Sun are 
$(g^\prime-r^\prime)_\odot$=0.44 and
$(r^\prime-i^\prime)_\odot$=0.11. The measured colour index
$(r^\prime-i^\prime)$ is therefore consistent with the slightly  
blue colour respect to the Sun observed in the spectrum. In
    addition, we have obtained the colour index $(r^\prime-i^\prime)$
    from the P/2021 A5 binned spectrum, resulting in
    $(r^\prime-i^\prime)$=0.09, which is within the error bars of the measured
    index from image photometry.

For modelling purposes, we take the geometric
albedo of these objects as $p_v$=0.106 for P/2019 A4
\citep[see][median value for 
  $X$-type asteroids]{2012ApJ...745....7M}, and $p_v$=0.07 for P/2021 A5
\citep[see][averaged value for $B$-types]{2016A&A...591A..14A}. The
density of these bodies were taken as 1850 kg m$^{-3}$ and 2380 kg m$^{-3}$  for
P/2019 A4 and P/2021 A5, respectively \citep{2012P&SS...73...98C}. The
corresponding linear phase coefficients are calculated from the
magnitude-phase relationship by \cite{1997SoSyR..31..219S} as 0.036
mag deg$^{-1}$ 
and 0.04 mag deg$^{-1}$ for P/2019 A4 and P/2021 A5, respectively.

In order to set a stringent upper limit to the asteroid
    radii, we converted the absolute magnitudes observed to diameters using the
    usual relationship from  \cite{2007Icar..190..250P}. We remark
    that these diameters are just upper limits that we will use later in the
    mode-ling procedure, given
    the strong contamination of dust around those objects.       
Absolute magnitudes in the $r^\prime$ filter are computed as $H_r^\prime =
r^\prime - 5\log (R \Delta) - \eta \alpha$, where $\eta$ is the linear
phase coefficient. Assuming a $(B-V)$ colour index approximately solar
(see Table~\ref{stars}), and $r^\prime=V-0.49(B-V)+0.11$ \citep{1996AJ....111.1748F}, we get
$V=r^\prime+0.21$. Then, an estimate to the upper limits to the  
asteroids diameters can be made through the expression $D=\frac{1329}{\sqrt
  p_v} 10^{-0.2 H_v}$ \citep{2007Icar..190..250P}, resulting in 
$D$=0.7-1.0 km for P/2019 A4, and $D$=2.4 km for P/2021 A5.

\section{Dust modelling}

We proceed in the same way as in previous works, were we made use of our Monte
Carlo dust tail code to characterise the dust environment of comets
and active asteroids. Recent applications of the model can be found in
\cite{2019A&A...624L..14M} and \cite{2020MNRAS.495.2053D}. The idea is
to build up a dust tail for a certain object at a given time given a
series of input parameters such as the size distribution of the particles,
dust mass loss rate, and ejection speeds. The particles are assumed to
be characterised by their density, the geometric albedo, and the
linear phase coefficient. In the Monte Carlo procedure, the tail
brightness is computed by adding up the
contribution to the brightness of each particle ejected from the comet
or asteroid nucleus. The trajectory of each particle depends on 
the ratio of radiation pressure force to the gravity force, i.e., the
$\beta$ parameter, which is defined as  $\beta =C_{pr}Q_{pr}/(2\rho r)$,  where
$C_{pr}$=1.19$\times$ 10$^{-3}$ kg m$^{-2}$ is the radiation
pressure coefficient, $Q_{pr}$ is the scattering efficiency for
radiation pressure, and $\rho$ is the particle density. $Q_{pr}$ 
is taken as 1, as it converges to that value for absorbing particles
of radius $r \gtrsim$1 $\mu$m  
\citep[see e.g.][their Figure 5]{2012ApJ...752..136M}. Since the
number of physical parameters is large, several assumptions must be
made to make the problem tractable. We assume the geometric
albedo, linear phase coefficient, and density, as estimated in the
previous section for each asteroid. In addition, the particles are assumed to be
distributed in a broad power-law size distribution function $n(r) \propto
r^\kappa$, where $\kappa$ is the power-law index.

If the dust production rate is not sufficiently high as to dominate
the scattering cross section, the nucleus surface might have a
significant contribution to 
the observed brightness. For the  two targets under study, this might
be the case, so we have added up the nucleus contribution to the
brightness by considering its effective cross section and the 
geometric albedo and linear phase coefficient given in the previous
section. The nuclear radii ($R_N$) are then 
considered as one of the free parameters of the model, always
subjected to the constraint of being smaller than the upper limits
estimated in Section 2 from the absolute magnitudes. 

The remaining dust parameters involved in the dust tail brightness
computation are taken as free parameters to be fitted by a
multidimensional fitting algorithm, namely the downhill simplex method
of \cite{NelderMead65}, which has been implemented in \texttt{FORTRAN} language 
by \cite{1992nrfa.book.....P}. The best-fit parameters are found by
minimising the squared sum of the differences between the modelled and
measured tail brightness for the GTC images. To perform an appropriate
comparison between observed and modelled 
tails, each modelled tail is convoluted with a Gaussian function
having a \texttt{FWHM}  
equal to the measured seeing. The fitting parameters are the
power-law exponent of the size distribution $\kappa$, the ejection speeds, and
the dust mass loss rate. To keep the number of free parameters to a
minimum, the dust mass loss rate function is assumed as a Gaussian
function with parameters $M_t$ (the total dust mass ejected), the
time of maximum dust loss rate ($t_0$) and the 
\texttt{FWHM} of the Gaussian, which gives a measure of the
effective time span of the emission event. The peak dust loss rate,
$(dM/dt)_0$, is 
related to the total dust mass loss and the \texttt{FWHM} through the
equation  $M_t=1.06(dM/dt)_0\texttt{FWHM}$. The ejection speed is
assumed to follow the equation  
$v=v_0\beta^{\gamma}$, where $v_0$ and $\gamma$ are fitting
parameters of the model. Then, the set of seven fitting parameters ($NP=7$) is
$M_t$, $t_0$, \texttt{FWHM}, $v_0$, $\gamma$, $\kappa$, and the nucleus
radius, $R_N$. To
start the execution of the code, an initial simplex must be set
with $NP$+1=8 sets of input parameters that we choose to vary broadly
between reasonably expected minimum and maximum limits. Since the
best-fit set of parameters found in the downhill simplex method
necessarily corresponds to a
local minimum of the fitting function, we repeated the procedure for a
variety of input starting simplex 
parameters in an attempt to find the lowest of those local minima.

\begin{figure}
\centerline{\includegraphics[angle=-90,width=\columnwidth]{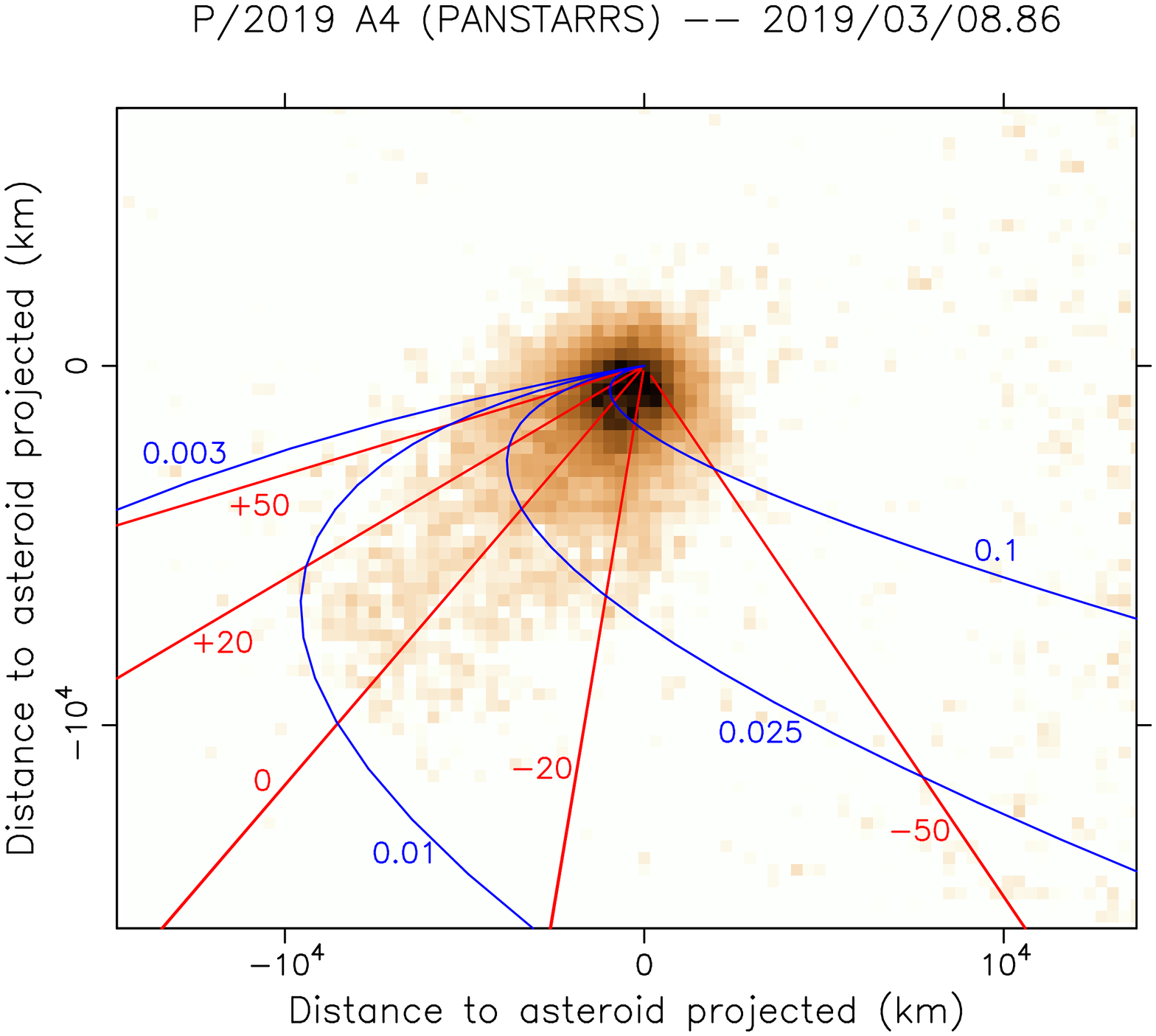}}
\caption{Syndyne-synchrone network for asteroid P/2014 A4 on 2019 March
  8.86. Synchrones (in red) are labelled in days since perihelion
  passage. Syndynes (in blue) are labelled in cm.}
   \label{synchroA4}
\end{figure}

\begin{figure}
\centerline{\includegraphics[angle=-90,width=\columnwidth]{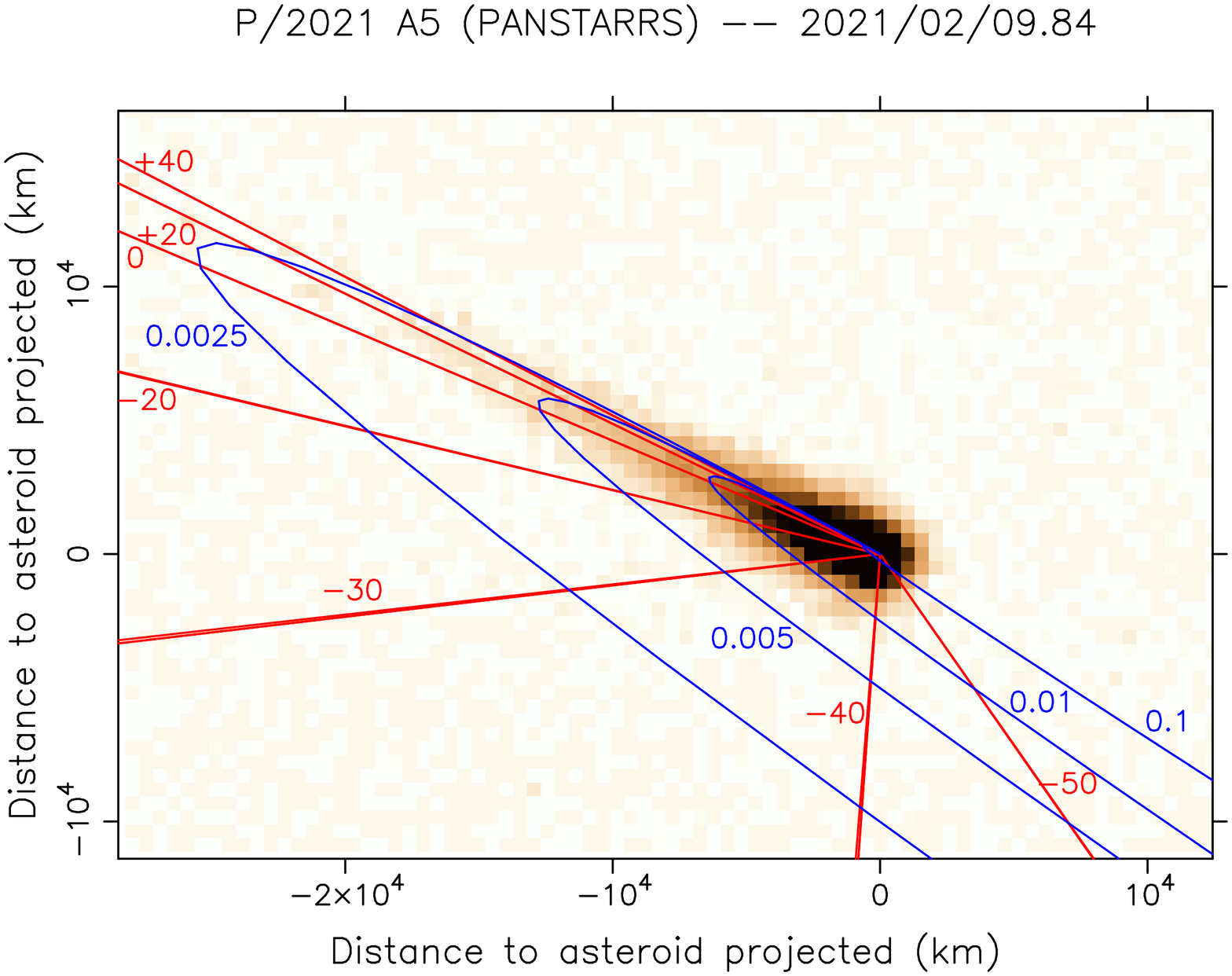}}
\caption{Syndyne-synchrone network for asteroid P/2021 A5 on 2021 February 9.84.
  Synchrones (in red) are labelled in days since perihelion
  passage. Syndynes (in blue) are labelled in cm.}
   \label{synchroA5}
\end{figure}

Some information on the ejection event timings and the size of the
ejected particles can be retrieved 
from the syndyne-synchrone network associated to each observation
date. For asteroid P/2019 A4, we show the syndyne-synchrone map
associated to the image taken on 2019 March (see Figure~\ref{synchroA4}). This network suggests that
dust emission must have occurred near perihelion, and likely concentrated
$\sim\pm$20 days respect to that date. In addition, particles smaller
than about 30 $\mu$m do not contribute significantly to the tail brightness,
although one must always keep in mind that the syndyne-synchrone map
refer to the geometric loci of particles ejected with zero velocity from the nucleus. For
P/2021 A5, the syndyne-synchrone map for the observation date is
plotted 
in Figure~\ref{synchroA5}. In this case, we see that the activity must have occurred 
when the object was approaching perihelion and later. Owing to the
geometry of the observation from Earth, post-perihelion synchrones
are closely spaced becoming more difficult in this
case to predict an event date from the syndyne-synchrone network. The
syndyne curves indicate 
that dust particles smaller than about 30 $\mu$m contribute very little to
the observed tail.

We use these syndyne-synchrone networks to establish safe limits in
the starting simplex for
the time interval during which the asteroids were active, in terms of
$t_0$ and \texttt{FWHM} of the Gaussian defining the dust loss rate profile, as well as 
the minimum and maximum particle radii of the size distribution
function, which we set to 10 $\mu$m and 1 cm, respectively, for both asteroids. 

\begin{figure*}
\centerline{\includegraphics[angle=-90,width=\textwidth]{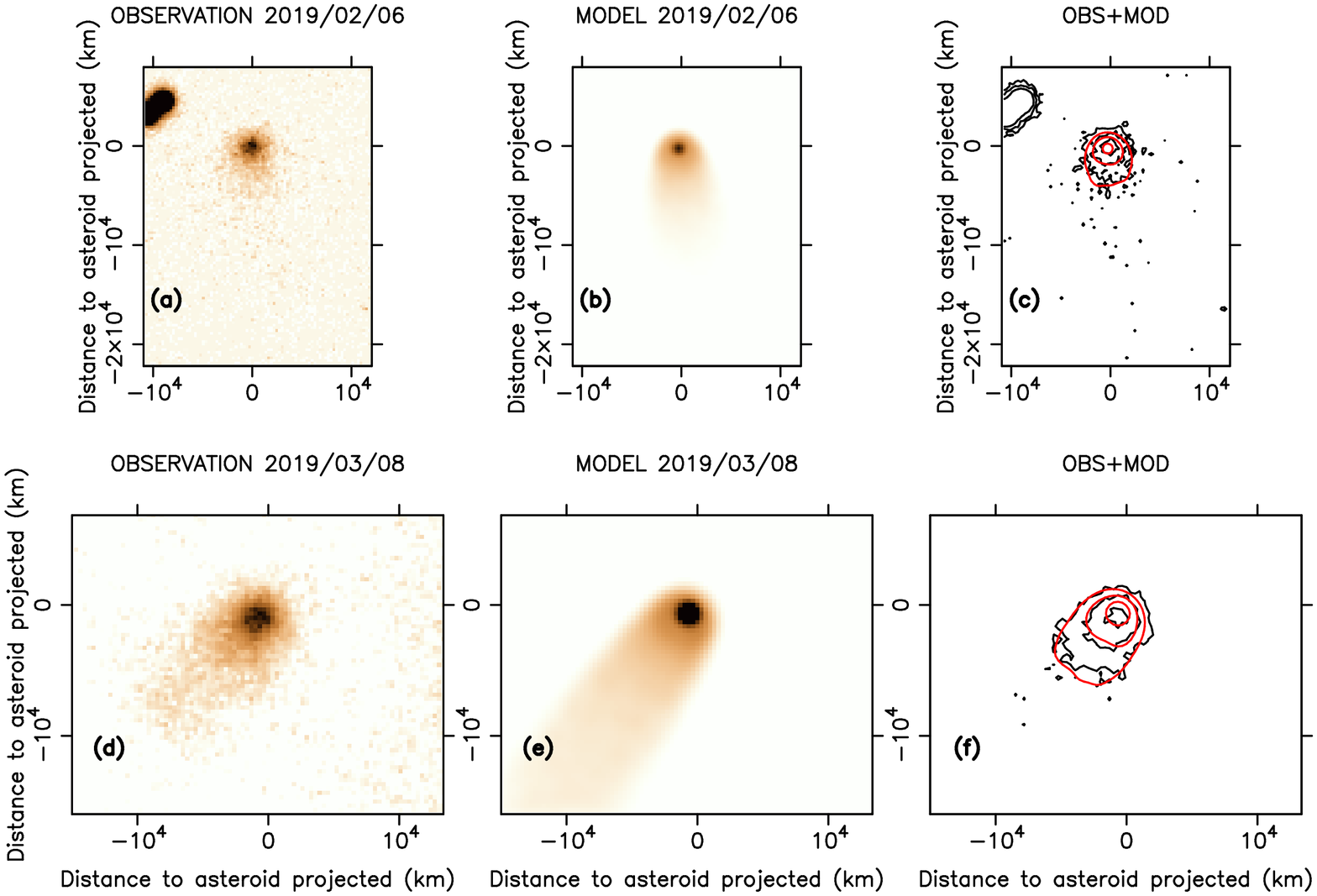}}
\caption{Observation and best fit model for P/2019 A4. The best-fit
  model parameters are the nominal parameters in Table~\ref{bestfit}. Upper panels:
  observation (a) and model (b) images on 2019 February. The rightmost
  panel (c) displays the observed (black contours) and modelled (red
  contours) isophotes. Innermost isophote corresponds to
  2$\times$10$^{-14}$ solar disk intensity units. Isophotes decrease in
  factors of two outwards. Lower panels: same as the upper panels but for
  the 2019 March image. Isophotes decrease in factors of two
  outwards, the innermost one corresponds to 5$\times$10$^{-15}$ solar
  disk intensity units. 
}
   \label{fitA4}
\end{figure*}

\begin{figure}
\centerline{\includegraphics[angle=-90,width=\columnwidth]{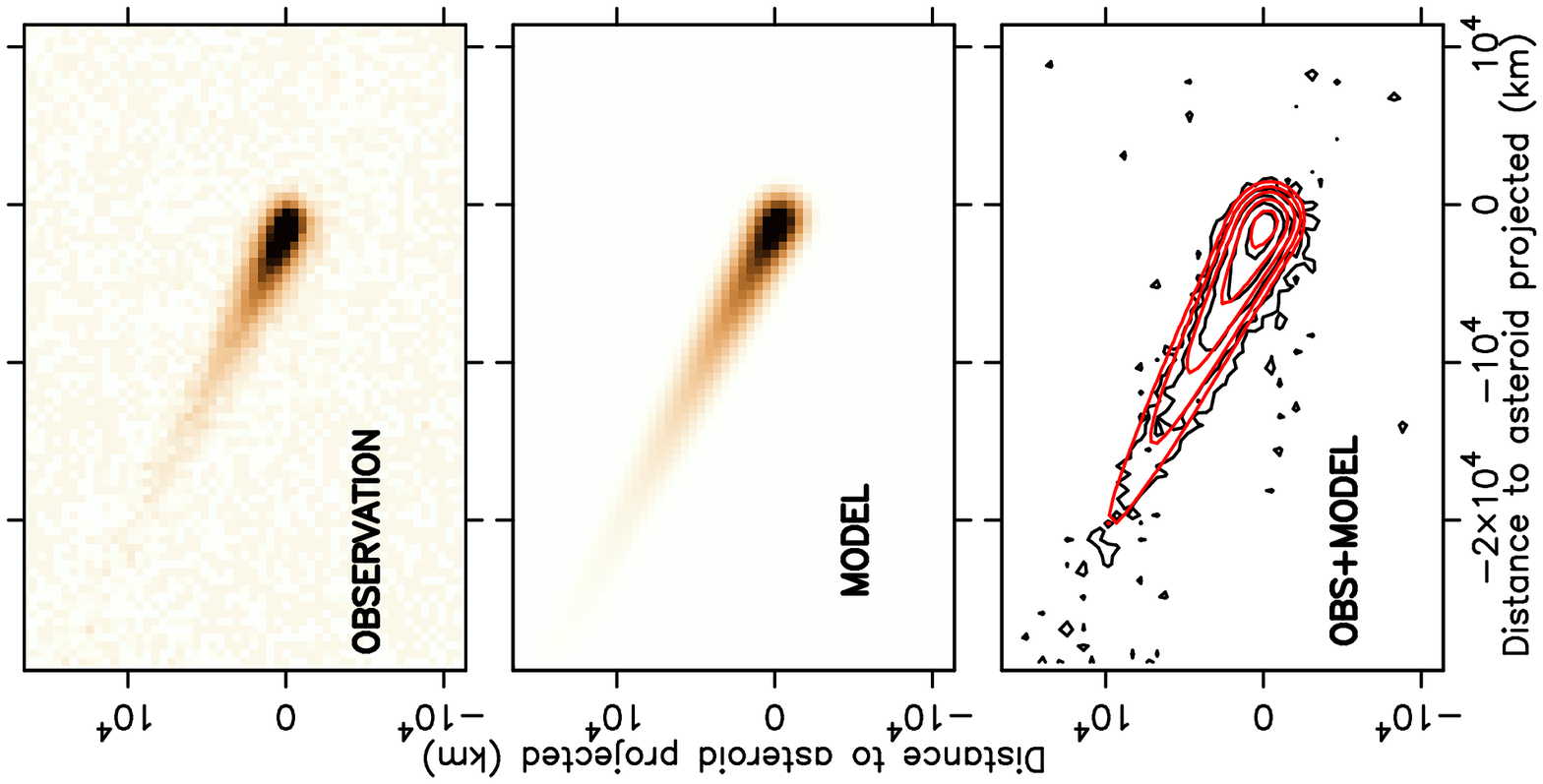}}
\caption{Observation and best fit model for P/2021 A5, as
  indicated. The best-fit
  model parameters are the nominal parameters in Table~\ref{bestfit}. In
  the lowermost panel, the observed (black contours) and modelled (red
  contours) isophotes are displayed. Innermost isophote corresponds to
  3$\times$10$^{-14}$ solar disk intensity units. Isophotes decrease in
  factors of two outwards. 
}
   \label{fitA5}
\end{figure}

\section{Results and discussion for P/2019 A4}

For asteroid P/2019 A4 we found the best-fit parameters as shown in 
Table~\ref{bestfit}, and the observed and simulated tails in Figure~\ref{fitA4}. The \texttt{FWHM} is
always constrained to \texttt{FWHM}$<$20 days, and the total dust
mass loss is given by $M_t$=($2.0\pm$0.7)$\times$10$^6$ kg.
This short duration event indicates that the most  
probable cause is either an impact or a rotational
disruption. Although the time of peak activity is very close to
asteroid perihelion, we rule out, in principle, ice sublimation as a
possible mechanism 
because of the asteroid position in the central part of the belt, a 
too hot location for stable reservoirs of water ice to exist,
and its orbital stability over a timescale of 100 Myr or
    longer. In
addition, the low eccentricity of this object makes ice sublimation at
perihelion even less likely.

The
nominal ejection speeds ($v=1.6 \beta^{0.2}$ m s$^{-1}$) are found to depend
slightly on the particle 
radius (the exponent, $\gamma$, is $\gamma$=0.2), ranging from a minimum of 0.2 m s$^{-1}$
to a maximum of 0.8 m s$^{-1}$ for particles of radii 1 cm and
10 $\mu$m, respectively. The derived nominal nucleus radius, $R_N$=170 m,
implies an escape speed of $v_{esc}$=0.23 m s$^{-1}$ which is very 
consistent with the ejection speed of the largest particles considered ($r$=1
cm). 

If an impact were the cause of the observed activity, the impactor's
size can be roughly estimated from the ejected mass following an argument
similar to that by \cite{2013ApJ...764L...5J} for asteroid P/2010
A2. For an average collision speed of $\sim$5 km s$^{-1}$ in the main
belt, and an ejecta escape velocity of 0.23 m s$^{-1}$, the ratio of
ejecta mass to projectile mass, assuming that impactor and target have
the same density, is of order $M_e/M_p\sim$10$^4$
\citep{2011Icar..211..856H}. Then, if the ejected mass is $M_e=2\times 10^6$
kg, the impactor mass becomes $M_p$=200 kg, corresponding to a
spherical object of only 0.3 m in radius ($\rho$=1850 kg m$^{-3}$). On
the other hand, a rotational disruption might have also occurred: the
low speed of the particles ejected and the fact that the 
event might last as much as 20 days point to that
possibility. Besides, the timescale for
Yarkovsky-O'Keefe-Radzievskii-Paddack (YORP) induced rotational
acceleration for objects having radius $\lesssim$6 km 
is always shorter than the collision timescale
\citep{2014MNRAS.439L..95J}. Thus, for an asteroid at 2.5 au from the
Sun and a YORP coefficient of
$Y$=0.01 (which is related to the 
asteroid shape), $\tau_{YORP}=42 R_A^2$ Myr, where $R_A$ is the radius
of the asteroid in $km$ \citep[see][]{2014MNRAS.439L..95J}, resulting
in $\tau_{YORP}$=1.2 Myr for P/2019 A4. The
collisional timescale for the nominal asteroid radius (170 m) is
$\tau_{coll}\sim$140 Myr \citep[see][their Fig. 14]{2005Icar..179...63B}, i.e., more than
two orders of magnitude longer. However, at this point we are unable
to favour one of these two mechanisms. Observations of the lightcurve
of the target could help in solving the problem. Nevertheless, the
faintness of the object might preclude any future attempt to measure
its rotation period.


\begin{table*}
	\centering
	\caption{Nominal best-fit parameters (values between brackets)
          and possible range of parameters for asteroids P/2019 A4 and
          P/2021 A5.}
	\label{bestfit}
	\begin{tabular}{lccccccc} 
	  \hline
Asteroid &  Total ejected & Time of max. activity & \texttt{FWHM} & $v_0$ & $\gamma$ & $\kappa$ &
$R_N$\\
   & dust mass ($M_t$, kg) & ($t_0$, days since perihelion) & (days) & (m s$^{-1}$) &  & & (m) \\ 
		\hline
P/2019 A4 & (2.0$\pm$0.7)$\times$10$^6$ [2.0$\times$10$^6$] & --3$\pm$10
[--3] &  $<$20 [7] &
1.6$\pm$0.4 [1.6] & 0.2$\pm$0.1 [0.2] & --3.2$\pm$0.1 [--3.2] &
170$\pm$70 [170] \\
P/2021 A5 & (8.0$\pm$2.0)$\times$10$^6$ [8$\times$10$^6$] & +48$\pm$30
[+48] & 5-60 [39] & $<$0.3 [0.15]
& <0.1 [0.0] & --3.4$\pm$0.1 [--3.4] & <500 [150] \\ 
		\hline
	\end{tabular}
\end{table*}

\section{Results and discussion for P/2021 A5}

Table~\ref{bestfit} gives the results for the best-fit parameters concerning
asteroid P/2021 A5, and the observed and modelled images in
Figure~\ref{fitA5}.
In this case, the event time is shifted by +48
days (nominally)
after perihelion passage, but with a large uncertainty of $\pm$30
days. The event duration is constrained between 5 and 60
days, and the total dust mass loss released is
$M_t$=(8$\pm$2)$\times$10$^6$ kg. The nominal particle 
ejection speeds (0.15 
m s$^{-1}$) are found to be independent of size ($\gamma$=0.0). The escape speed
corresponding to the derived $R_N$=150 m nucleus is 0.17 m s$^{-1}$,
which is of the same order of the ejection speeds.  Regarding the
activation mechanism(s), the orbital dynamics of a few percent of the dynamical clones  
point to a JFC origin for this object and, consequently, it might be
an 
ice-bearing asteroid. The ice-driven activity is compatible with the duration of the activity,
\texttt{FWHM}=39 days (nominal). However, mass shedding from rotational
instabilities is also feasible. Perhaps the less probable cause is an
impact, because of the lower limit of the duration of \texttt{FWHM}$>$5 days,
although it cannot be ruled out. Further observations of this target
in future apparitions could shed light on the responsible
mechanism(s).

\section{Conclusions}

Key conclusions on the observation and modelling of active asteroids
P/2019 A4 and P/2021 A5 are as follows.

1) Orbital dynamics simulations show that P/2019 A4 moves in a
    stable orbit over long (100 Myr) timescales. Asteroid P/2021 A5
   is located close to the 9:4 Jupiter resonance region and the
   dynamical evolution of 4\% of its clones indicates a possible JFC origin.

2) The spectrum of P/2019 A4 is slightly red with slope $S'=4 \pm 1
\%/100$ nm typical of a $X$-type asteroid while the 
of P/2021 A5 spectrum is slightly blue with a slope $S'= -3 \pm 1
\%/100$ nm, typical of a primitive asteroid of $B$ type.

3) From the Monte Carlo dust tail modelling of GTC images on these targets, we
estimate that P/2019 A4 was active for a maximum period of 20 days
(\texttt{FWHM}), which indicates, probably, an activation mechanism related to
either a rotational disruption or an impact. This is also supported by
the stability of its orbit in the central part of the belt. This increases the statistics of
bodies in the middle belt whose 
activity is reported as short-lived. Regarding P/2021 A5, the longer possible
activity period up to 
60 days, and the possibility of a JFC origin as revealed by the
orbital dynamics simulations could indicate an
ice-driven activity, but we are unable to 
rule out other hypotheses due to the large
uncertainty (plausible \texttt{FWHM} values in the 5-60 days range) in the
estimated duration. 

4) The total dust mass ejected from these two asteroids are, for
a maximum particle radius ejected of 1 cm,  
(2.0$\pm$0.7)$\times$10$^6$ kg and (8$\pm$2)$\times$10$^6$ kg for
P/2019 A4 and P/2021 A5, respectively, and the ejection is
concentrated close to the perihelion passage for P/2019 A4, and
possibly shortly after perihelion passage for P/2021 A5, although with
an uncertainty of $\pm$30 days. 

5) The derived ejection speeds for both targets $\sim$0.2 m s$^{-1}$
are consistent with the escape speeds of the nuclear radii estimated 
from the Monte Carlo modelling, which turned out to be in 
the range of 100-240 m (nominal value 170 m) for P/2019 A4, and
smaller than 500 m (nominal value 150 m) for P/2021 A5.

\section*{DATA AVAILABILITY}

The spectral and image data shown in this manuscript are available
from the GTC Public Archive after the one-year proprietary period is over.

\section*{Acknowledgements}

We are deeply indebted to an anonymous referee who, among other
constructive and useful
suggestions, draw our attention to the possibility that asteroid P/2021 A5
might be a JFC interloper, and encouraged us to carry out the
corresponding orbital dynamical calculations that have indeed indicated that
this might be the case.   

This work is based on observations made with the Gran Telescopio Canarias (GTC),
installed at the Spanish Observatorio del Roque de los Muchachos of
the Instituto de Astrof\'\i sica de Canarias, in the island of La Palma.

FM and DG acknowledge financial support from the State Agency for Research of
the Spanish MCIU through the "Center of Excellence Severo Ochoa" award
to the Instituto de Astrof\'\i sica de Andaluc\'\i a (SEV-2017-0709).
FM and DG also acknowledge financial support from the  Spanish  Plan
Nacional  de  Astronom\'\i a  y  Astrof\'\i sica  LEONIDAS
project RTI2018-095330-B-100,  and  project  P18-RT-1854  from  Junta
de  Andaluc\'\i a.

\bibliographystyle{mnras}
\bibliography{papermnras}

\bsp	
\label{lastpage}
\end{document}